\documentclass[         %
aps,                    
prd,                    
showpacs,               
nofootinbib,            
twocolumn,             %
showkeys,               %
preprintnumbers,        %
floatfix]               
{revtex4}               
\usepackage{graphicx,longtable} 

\begin{document}
\title{Constraints on Axial Two-Body Currents from Solar Neutrino 
Data}
\author{        A.~B. Balantekin}
\email{         baha@nucth.physics.wisc.edu}
\author{        H. Y\"{u}ksel}
\email{         yuksel@nucth.physics.wisc.edu}
\affiliation{  Department of Physics, University of Wisconsin\\
               Madison, Wisconsin 53706 USA }
\date{\today}
\begin{abstract}
We briefly review recent calculations of neutrino deuteron  
cross sections within the effective field theory and traditional  
potential model approaches. We summarize recent efforts to 
determine the counter term describing axial two-body currents, 
$ L_{1A} $, in the effective field theory approach. 
We determine the counter term directly from the 
solar neutrino data and find several, slightly different, ranges of 
$ L_{1A} $ under different sets of 
assumptions. Our most conservative fit value with the largest 
uncertainty is 
$ L_{1A} = 4.5 ^{+18}_{-12} \: \mathrm{fm}^3$.  
We show that the contribution of the uncertainty of $ L_{1A} $ to 
the analysis and interpretation of the solar neutrino data measured 
at the Sudbury Neutrino Observatory is significantly less than the 
uncertainty coming from the lack of having a better knowledge 
of $ \theta_{13} $. 
\end{abstract}
\medskip
\preprint{} 
\maketitle


A significant amount of theoretical work was recently directed 
towards the calculation of neutrino capture on deuteron. Some of 
these efforts to describe this process utilized the effective 
field theory approach. In this approach nonlocal interactions at 
short distances are represented by effective local interactions 
in a derivative expansion. Since the effect of a given operator on 
low-energy physics is inversely proportional to its dimension,  
an effective theory valid at low energies can be 
written down by retaining operators up to a given dimension. The 
coefficients of these operators are then needed to be fixed either 
directly by the data or can be fitted to the results of 
calculations carried out using more traditional approaches. 

For nucleon-nucleon interactions it was shown that one can 
introduce a well-defined power counting \cite{Kaplan:1998tg}. 
In this method one needs to introduce a single coefficient, 
commonly called $ L_{1A} $, to parameterize the unknown 
isovector axial two-body current which dominates the 
uncertainties of all neutrino-deuteron interactions. Using an  
effective theory without pions \cite{Kaplan:1998sz} such a 
calculation was carried out in Ref. \cite{Butler:1999sv}. These 
authors found that the ratio of charged- to neutral-current was  
fairly insensitive to this counter term. To test the convergence 
of the results in Ref. \cite{Butler:1999sv} Butler, Chen, and 
Kong also calculated next-order corrections and found that no new
parameters need to be introduced \cite{Butler:2000zp}. An 
alternative formulation of the effective field theory approach 
using heavy-baryon chiral perturbation theory was given in 
Ref. \cite{Ando:2002pv}.

The cross section for neutrino absorption on deuterium was first 
calculated in Refs. \cite{Kelly:gy} and \cite{BahcEll} utilizing 
an effective range approximation to describe the nuclear 
interaction, using the allowed approximation 
for the weak operators, and assuming that the final two-nucleon 
state has a relative angular momentum of zero. First-forbidden 
contributions to the weak operators were included using 
Siegert's theorem in Ref. \cite{Ying:1989as} and using convection 
current form of the vector operators in Ref. \cite{Tatara:eb}. 
A detailed assessment of various approximations in these papers 
was given in Refs. \cite{Ying:1991tf} and \cite{Doi:tm} using 
various nuclear potentials. This work was recently updated in 
Ref. \cite{Nakamura:2000vp}. 

Radiative corrections to the charged-current breakup of the 
deuteron were calculated by Towner in Ref. \cite{Towner:bh}. 
Beacom and Parke pointed out \cite{Beacom:2001hr} inconsistencies 
in Towner's treatment of radiative corrections.
This inconsistency was  
resolved in Ref. \cite{Kurylov:2001av} and cross section  
calculations using more recent values of $ g_A $ were given in Refs. 
\cite{Kurylov:2001av} and \cite{Nakamura:2002jg}. More recently it  
was shown that radiative corrections to the charged-current  
neutrino-nuclear reactions with either an electron or a positron 
in the final state are described by a universal function  
\cite{Kurylov:2002vj}. 

The counter term, $ L_{1A} $, describing the effects of the leading 
weak axial two-body current can be determined either by comparing 
various cross sections calculated using the effective field theory 
approach with those calculated using standard potential model 
approach or with experimentally or observationally determined cross 
sections. When the renormalization scale is set to the muon mass, 
dimensional analysis gives a rough estimate of this quantity 
\cite{Butler:2000zp}
\begin{equation}
\vert L_{1A} \vert \sim 6 \: \mathrm{fm}^3 .
\end{equation} 
It should be emphasized that 
this number depends on the renormalization scale and cannot be 
reliably used at lower energies. Using the existing reactor  
antineutrino-deuteron breakup data provides a constraint of 
\cite{Butler:2002cw} 
\begin{equation}
L_{1A}  = 3.6 \pm 5.5  \: \mathrm{fm}^3 .
\end{equation} 

\begin{figure*}[t]
\vspace*{+.2cm}
\includegraphics[scale=.6]{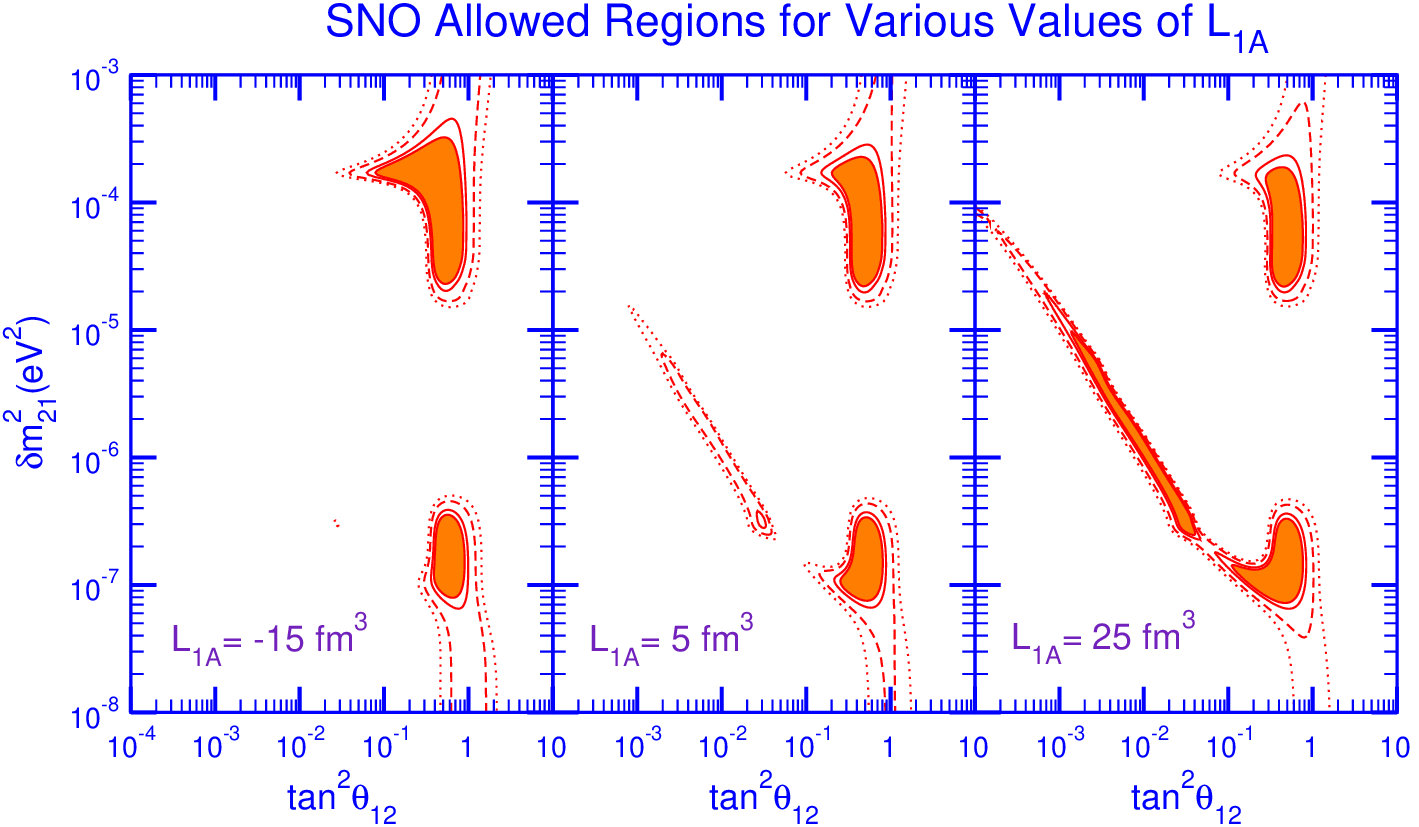}
\vspace*{+.0cm} \caption{ \label{fig:1} 
The change in the allowed region of the neutrino parameter space 
using solar neutrino data measured at SNO as the value of 
$ L_{1A} $ changes. In the calculations leading 
to this figure the neutrino mixing angle $\theta_{13}$ is taken 
to be zero (see text). The shaded
areas are the 90 \% confidence level region. 95 \% (solid line),
99 \% (log-dashed line), and 99.73 \% (dotted-line) confidence
levels are also shown.} 
\end{figure*}

Helioseismic observation of the pressure-mode oscillations of the 
Sun can be used to put constraints on various inputs into the 
Standard Solar Model, in particular the $pp$ fusion cross section. 
This process has been calculated to the 
fifth order in pionless effective field theory 
\cite{Butler:2001jj}. Neutrino-deuteron and antineutrino-deuteron  
scattering are computed to the third order in the same approach. 
The value of $ L_{1A} $ is not the same in different orders. 
Helioseismology limits $ L_{1A}= 7.0 \pm 5.9  \: \mathrm{fm}^3$  
in the fifth order \cite{Brown:2002ih}. Using the expressions given 
in Ref. \cite{Butler:2002cw} this gives 
\begin{equation}
L_{1A}  = 4.8 \pm 6.7  \: \mathrm{fm}^3 
\end{equation}
in the third order. A state of the art calculation of the $pp$  
fusion cross section was given in \cite{Schiavilla:1998je} 
where the uncertainty in the axial two-body current operator was  
adjusted to reproduce the measured Gamow-Teller matrix element 
of tritium $\beta$ decay. After performing the transformation 
from the fifth- to third-order, the calculation of Ref. 
\cite{Schiavilla:1998je} indicates a value of
\begin{equation}
L_{1A}  = 4.2 \pm 2.4   \: \mathrm{fm}^3.
\end{equation}

One can also try to determine the counter term directly using the 
solar neutrino data. Using the Sudbury Neutrino Observatory (SNO)  
and SuperKamiokande (SK) charged-current, neutral current, and 
elastic scattering rate data Chen, Heeger, and Robertson (CHR) 
find \cite{Chen:2002pv}
\begin{equation}
L_{1A}  = 4.0 \pm 6.3 \: \mathrm{fm}^3 .
\end{equation}
In order to obtain this result CHR wrote the observed 
rate in terms of an averaged effective cross section and a 
suitably defined response function. In this paper we explore 
the phenomenology associated with the variation of $L_{1A} $.

In our calculations we used the neutrino cross sections given in 
Refs. \cite{Butler:1999sv} and \cite{Butler:2000zp}. The radiative 
corrections are taken into account following Ref.  
\cite{Kurylov:2001av}. To calculate observed solar neutrino rates 
and spectra we used a covariance approach the details of which are
described in Ref. \cite{Balantekin:2003dc}. In all calculations to
obtain the MSW survival probabilities we  
used the neutrino spectra and solar electron density profile 
given by
the Standard Solar Model of Bahcall and collaborators
\cite{Bahcall:2000nu}.

The dependence of the extracted neutrino parameters on the value of 
$L_{1A} $ is not very strong. We show how the parameter space 
changes with $L_{1A} $ in Figure 1. In this figure to find the 
allowed regions we fit 
34 data points from the SNO day-night-spectrum \cite{Ahmad:2001an} 
using the procedure of Ref. \cite{Balantekin:2003dc}. 
The shaded  
area is the 90 \% confidence level region. 95 \% (solid line), 
99 \% (log-dashed line), and 99.73 \% (dotted-line) confidence  
levels are also shown. As 
$L_{1A} $ changes from $-15 \: \mathrm{fm}^3$ to $25 \: 
\mathrm{fm}^3 $ we note that the changes in the shape of the 
confidence level intervals 
are small. In the calculations leading to this figure we took 
the total $ ^8{\mathrm{B}} $ flux to be a free parameter using the 
procedure discussed in 
Ref. \cite{Balantekin:2003dc}. (Note that even though 
we show the entire parameter space in this figure in the rest of 
this paper we concentrated in the large mixing angle region which 
is preferred by the global analysis). We 
conclude that 
the uncertainty in $L_{1A} $ can not be a significant source of 
error in the analysis of SNO data. 

In general, both the deuteron breakup cross section and the 
total count rate at SNO are nonlinear in $L_{1A} $. Since 
$L_{1A} $ is small the charged- and neutral current count rates
can be linearized by making a first order expansion, i.e.
\begin{equation}
\mathrm{Count} \, \mathrm{Rate} \sim A + B \, L_{1A}, 
\end{equation}  
as was done by CHR. 
Both the energy dependence and  
the overall magnitude of the $^8$B flux is an input into the 
Standard Solar Model. The energy dependence is rather accurately 
determined by the laboratory measurements of the $^8$B decay. The 
overall magnitude is determined by the measured rate 
of the $ ^7{\mathrm{Be}} (p,\gamma) ^8{\mathrm{B}} $ reaction 
(for a review see Ref. \cite{Adelberger:1998qm}). 
To account for the sensitivity of the calculations on the value of 
the $^8$B flux we set $\Phi (^8{\mathrm{B})} = f_B  
\Phi_{\mathrm{SSM}} (^8{\mathrm{B})}$ and calculate the total rate 
for various values of the parameter $ f_B $. Clearly the total 
count rate should be proportional to the value of $ f_B $. 
Note that the elastic scattering count rate 
is independent of $L_{1A} $. Thus allowing $ f_B $ vary freely 
cannot be fully 
compensated by changing $L_{1A} $ as we discuss below. 

\begin{figure}[t]
\vspace*{+.3cm}
\includegraphics[scale=.49]{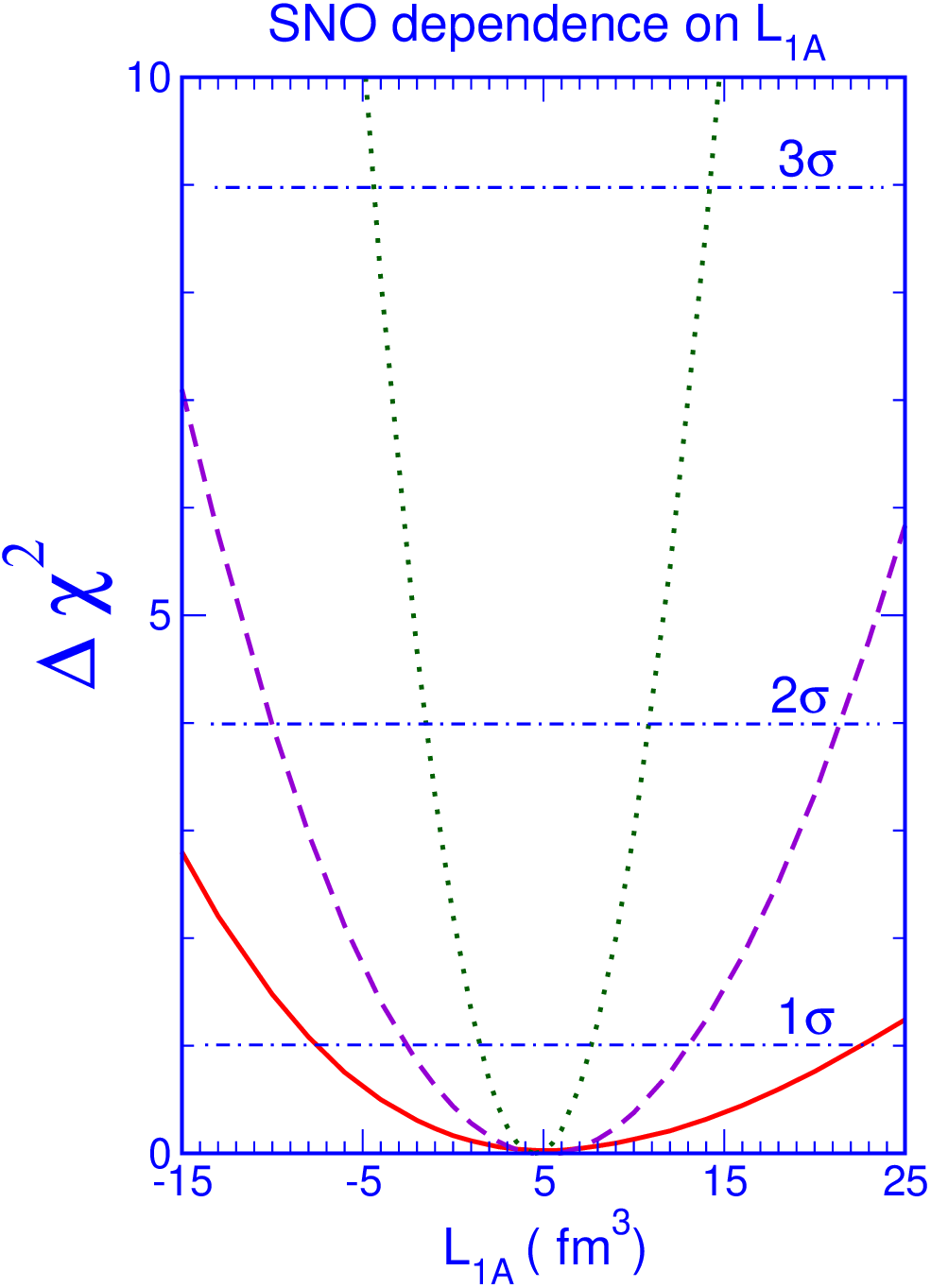}
\vspace*{+0cm} \caption{ \label{fig:2}
Projection of the global $\Delta \chi^2$ function
on the parameter $ L_{1A} $. In the calculations leading to this  
figure $\theta_{13}$ is taken to be zero. $\theta_{12}$, $ \delta 
m_{12}^2 $, and the parameter $f_B$ (the multiplier 
of the $^8$B flux) are varied. The solid  
line represents the case where all these three are unconstrained. 
The long-dashed line is when the $^8$B flux is fixed as described 
in the text, but the other two are unconstrained. 
The dotted line is when $\theta_{12}$ and $ \delta 
m_{12}^2 $ are also taken to be the best fit values to the SNO 
energy spectra.}  
\end{figure}

In Figure \ref{fig:2} we present the quantity 
$\Delta \chi^2 = \chi^2 - \chi^2_{\rm
min}$  calculated as a function of $L_{1A} $. In this figure
$\Delta \chi^2$ is projected only on one parameter ($ L_{1A}$) 
so that $n-\sigma$ bounds on it are given by  $\Delta \chi^2 = 
n^2$. $\theta_{13}$ is assumed to be zero. The solid line 
represents the case in which all other parameters ($\theta_{12}$, 
$ \delta m_{12}^2 $, and $f_B$) are unconstrained. The best fit  
value is given by $ L_{1A}  = 4.5 \: \mathrm{fm}^3 $. In this case 
$ L_{1A}$ is constrained between $- 7 \: \mathrm{fm}^3 $ and 
$ 23 \: \mathrm{fm}^3 $ at $ 1\sigma $ level. Such a wide range 
is not surprising since the dependence of the rate on $ L_{1A}$ is  
small and the effects of  
the parameters like $\theta_{12}$, $ \delta m_{12}^2 $, and $f_B$ 
are much more dominant. In order to obtain a better bound on 
$ L_{1A}$, we fix $f_B$ so that the total count rate of SNO  
\cite{Ahmad:2001an} is exactly reproduced at the value of 
$ L_{1A}$ which corresponds to the minimum $\chi^2$ of the fit 
while the parameters $\theta_{12}$ and $ \delta m_{12}^2 
$ are unconstrained. The resulting fit is shown by the dashed 
line. In this case $ L_{1A}$ is constrained between $- 2 \: 
\mathrm{fm}^3 $ and $ 13 \: \mathrm{fm}^3 $ at $ 1\sigma $ level. 
The dotted line in this figure represents the case where we fix 
all the parameters except $ L_{1A}$.
We find the best fit values of 
$\theta_{12}$ and $ \delta m_{12}^2 $ in a global fit using  
93 data points from solar and reactor neutrino experiments; namely 
the  total rate of the chlorine  
experiment (Homestake \cite{Cleveland:nv}), the average rate of 
the gallium experiments (SAGE \cite{Abdurashitov:2002nt}, GALLEX  
\cite{Hampel:1998xg}, GNO \cite{Altmann:2000ft}), 44 data 
points from the SK zenith-angle-spectrum \cite{Fukuda:2001nj}, 
34 data points from the SNO day-night-spectrum 
\cite{Ahmad:2001an} and 13 data points from the KamLAND spectrum  
\cite{Eguchi:2002dm}. In addition we fix $f_B$ so that the total 
count rate of SNO \cite{Ahmad:2001an} is exactly reproduced. 
If we were to exclude SNO data from the global analysis and took 
$f_B=1$ instead of fixing as described above this method would be 
tantamount to treating SNO as an 
experiment to measure only $ L_{1A}$ so that the uncertainties in 
the SNO data would only show up as the corresponding uncertainty 
at $L_{1A}$. From the dotted line $ L_{1A}$ is constrained between 
$ 2 \: \mathrm{fm}^3 $ and $ 8 \: \mathrm{fm}^3 $ at $ 1\sigma $ 
level. In a sense this latter range represents the "best case" 
limit on $ L_{1A}$ that one can obtain from SNO. It is worth 
emphasizing that the $ \chi^{2} $ minimum is almost the same in 
all these cases. In all cases we obtain a best fit value of 
$ L_{1A}$ around $4.5$ to $5$ $\mathrm{fm}^3 $ which is a little 
larger than the value obtained by CHR. These authors use elastic 
scattering, charged-current, and neutral-current rates separately
with effective cross sections. Since we fit the solar neutrino 
day-night spectrum directly by folding differential cross sections, 
detector response functions, $ ^  8{\mathrm{B}} $ spectrum, and the
MSW survival probabilities  
we need a slightly larger $ L_{1A}$.

\begin{figure*}[t]
\vspace*{+.0cm}
\includegraphics[scale=.61]{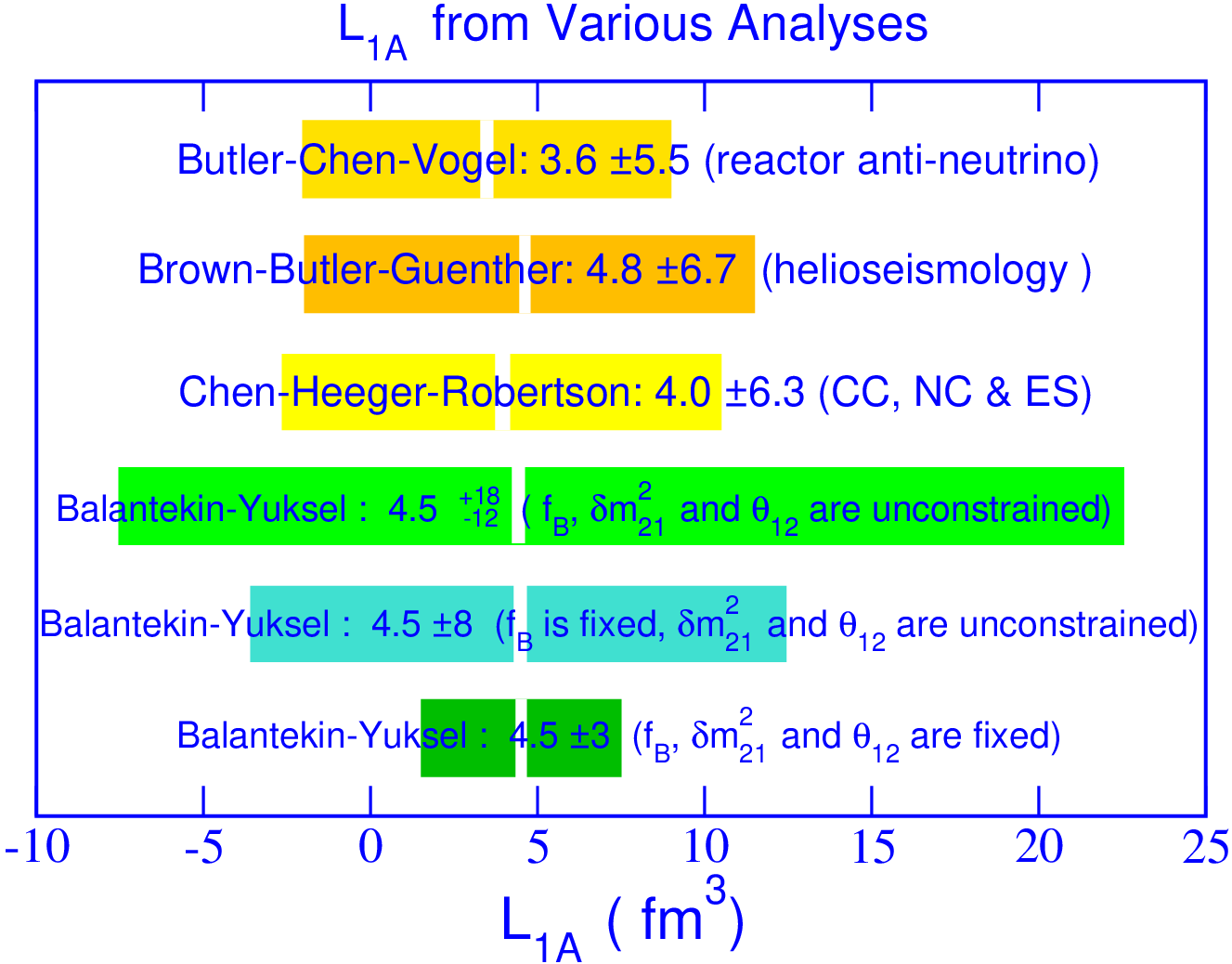}
\vspace*{+0cm} \caption{ \label{fig:4} 
Values of $ L_{1A} $ obtained from different analyses. The values 
labeled this work are calculated using the 
dashed and the dotted lines of 
Figure 2 as described in the text. Helioseismology limit is from 
Ref. \cite{Brown:2002ih}. Reactor antineutrino limit is from 
Ref. \cite{Butler:2002cw}. The limit obtained by CHS 
\cite{Chen:2002pv} is also shown.}  
\end{figure*}

In Figure \ref{fig:4} we compare our results with results from 
other analyses. Our results are based on the dashed and dotted 
lines of Figure \ref{fig:2}. We calculate $ 1\sigma $ errors by 
fitting a Gaussian of the form 
\begin{equation}
\exp \left[  - \frac{1}{2} \left( \frac{L_{1A} - 
L_{1A}^\mathrm{average}}{\sigma_{L_{1A}}} 
\right)^2 \right] 
\end{equation} 
to each side of the marginal likelihood expression $ \mathcal{L} 
= \exp \left( - \Delta \chi^2 /2 \right) $ 
and estimate two standard deviations separately for each side
\cite{bevrob}. 
We obtain the error band shown in the figure by 
symmetrizing those errors. 

One of the open questions in neutrino physics is understanding the  
role of mixing between the first and third flavor generations,  
$\theta_{13}$. In this regard we also explored if the 
uncertainties coming from the lack of knowledge of $\theta_{13}$ 
and the counter-term $L_{1A} $ are comparable. In the limiting case 
of small $ \cos {\theta_{13}}  $ and $ \delta m_{31}^2 
\gg \delta m_{21}^2 $, which seems to be satisfied by the 
measured neutrino properties, it is possible to incorporate the 
effects of $\theta_{13}$ rather easily. In this limit the 
three-flavor survival probability is 
is given by \cite{Kuo:1989qe,Fogli:2000bk,Balantekin:2003dc}

\begin{figure*}[t]
\vspace*{+.0cm}
\includegraphics[scale=.6]{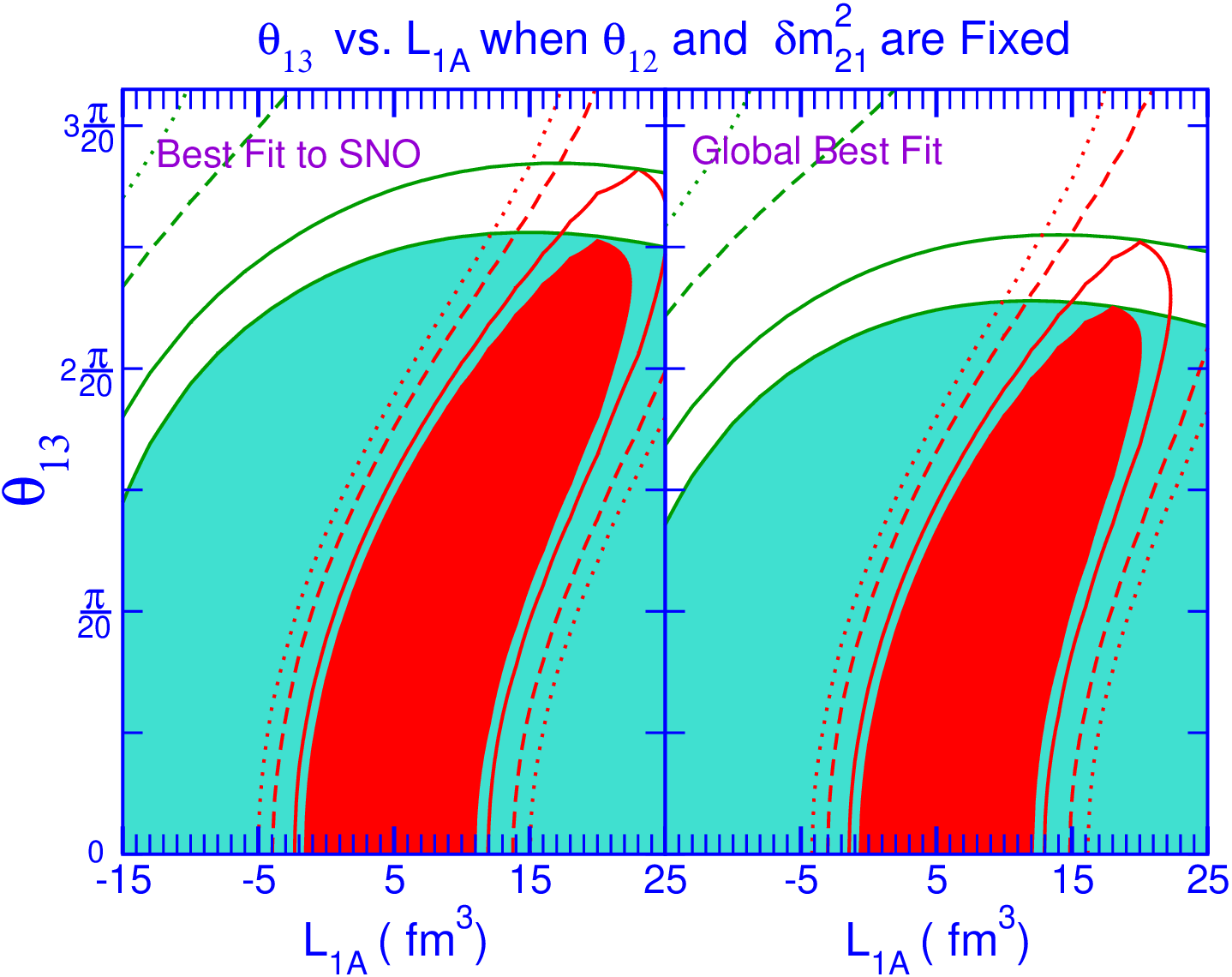}
\vspace*{+0cm} \caption{ \label{fig:5}
Allowed parameter space when $\theta_{12}$ and $ \delta 
m_{12}^2 $ are fixed to give the minimum $\chi^2$ values to 
reproduce the SNO day-night spectrum (left-hand panel) and all 
solar neutrino experiments along with the KamLAND experiment 
(right-hand panel). The shaded
areas are the 90 \% confidence level region. 95 \% (solid line),
99 \% (log-dashed line), and 99.73 \% (dotted-line) confidence
levels are also shown. The dark-shaded region corresponds to 
the case when the $^8$B flux is fixed to be the 
Standard Solar Model value. The light-shaded region corresponds 
to the case when that flux is unconstrained.}
\end{figure*}

\begin{eqnarray}
\label{threef}
P_{3\times3}( \nu_e \rightarrow  \nu_e) &=& \cos^4{\theta_{13}} 
\>P_{2\times2}( \nu_e \rightarrow  \nu_e \>{\rm with}\> N_e
\cos^2{\theta_{13}}) \nonumber \\ && + \sin^4{\theta_{13}}.
\end{eqnarray}

In Eq. (\ref{threef}) the quantity $P_{2\times2}(
\nu_e \rightarrow  \nu_e \>{\rm with}\> N_e \cos^2{\theta_{13}})$ is
the standard two-flavor survival probability calculated with the
modified electron density $N_e \cos^2{\theta_{13}}$ and the 
standard initial conditions. This suggests that for small values of 
$\theta_{13}$ the survival probability and consequently the counting  
rate can be linearized in $\cos^4{\theta_{13}}$: 
\begin{equation}
\mathrm{Count} \, \mathrm{Rate} \sim
A + B \, (1 - \cos^4{\theta_{13}} ).  
\end{equation}  
The neutral- and charged-current counting rates linearly depend on 
$L_{1A} $ while elastic scattering rate does not.  
Conversely the charged-current and elastic scattering rates 
linearly depend on $\cos^4{\theta_{13}}$ while the neutral-current 
rate does not. Hence it is 
reasonable to compare their relative contributions. To this end 
in Figure \ref{fig:5} we show the allowed $\theta_{13}$ 
and $L_{1A} $ parameter space when $\theta_{12}$ and $ \delta
m_{12}^2 $ are taken to give the minimum $\chi^2$ values to
reproduce the data. Results where the fixed values of 
$\theta_{12}$ and $ \delta
m_{12}^2 $  obtained 
using only the best fit of the SNO data (left-hand 
panel) and using the best fit of all the solar neutrino data along 
with KamLAND 
results (right-hand panel) are both shown. The dark-shaded 
region corresponds to the case when $f_B =1$. The light-shaded 
region corresponds to the case when the $^8$B flux is 
unconstrained. We observe that the 
uncertainty coming from the lack of a better 
knowledge of $\theta_{13}$ is larger than uncertainty coming from 
not knowing $L_{1A} $ precisely.  
One also observes that as $\theta_{13}$ increases the 
allowed $L_{1A} $  region shifts toward larger values of $L_{1A} $.
When $\theta_{13} = 0$ the electron neutrino flux is lost into only 
one channel: a particular linear combination of $ \mu $ and $ \tau
$ neutrinos \cite{Balantekin:1999dx}. But when $\theta_{13} \neq 0$ 
additional flux is lost into the orthogonal channel as well. The 
slightly decreased electron neutrino survival probability reduces 
the charged-current and the elastic scattering count rates so that 
a slightly larger $L_{1A} $ is needed to compensate the resulting 
decrease in the count rates at each bin. 

\newpage

In conclusion we showed that the SNO experiment with increased 
statistics using additional input from other solar neutrino 
experiments can significantly reduce the uncertainty in determining 
the precise form of the axial two-body currents at low energies. 
We also showed the contribution of the uncertainty in 
$L_{1A} $ to the analysis and interpretation of the SNO data is 
nearly negligible. The effect of this uncertainty is smaller than 
effects of a non-zero value of $\theta_{13}$ or even than effects  
of possible solar density fluctuations \cite{Balantekin:2003qm}. 
Finally our most conservative value for $L_{1A} $ is significantly 
larger than that was obtained by CHR. One reason for this may be the 
treatment of neutral- and charged-current count rates together 
in the global analysis.  


\section*{ACKNOWLEDGMENTS}
We thank R.G.H. Robertson and Jiunn-Wei Chen for valuable 
comments on the first version of the paper and 
M. Butler for useful discussions and providing us the 
cross sections of Refs. \cite{Butler:1999sv} and 
\cite{Butler:2000zp}.
This work was supported in part by the U.S. National Science
Foundation Grants No.\ PHY-0070161 and PHY-0244384 and in part by
the University of Wisconsin Research Committee with funds granted by
the Wisconsin Alumni Research Foundation.



\end{document}